# MCP-Enabled Agentic AI for Autonomous IPoDWDM Network Lifecycle Automation

Chunmin Xia, Jakub Harbaczewski, Nikhil Dsilva, Julie Raulin, Dominic Schneider, Achim Autenrieth

Adtran Networks SE, chunmin.xia@adtran.com

**Abstract** *This demo presents an MCP-enabled agentic AI architecture for autonomous control of vendor-agnostic IPoDWDM networks. We demonstrate live end-to-end lifecycle multi-layer automation and closed-loop control using GNPy and telemetry, validated on a real testbed. ©2026 The Authors*

## Overview

Increasing demand from AI workloads and high-bandwidth services is reshaping optical networks from traditional layered transport models and embedded optics toward next generation converged IPoDWDM infrastructures, enabled by high-speed single-channel rates such as 400G/800G coherent pluggables. The state of the art of IPoDWDM orchestrator uses a hierarchical controller to synchronize IP and optical SDN controllers[1,2]. However, this approach becomes complicated and hence expensive when multi-vendors located in different areas from both IP and optical networks are involved. In particular, achieving full automation for IPoDWDM control introduces significant complexity, as it requires tight vertical integration across network layers as well as horizontal integration across multiple vendors and heterogeneous control systems, posing challenges for cost-effective implementation.

In parallel, the explosive growth of agentic AI is widely regarded as a potential breakthrough for achieving universal and scalable automation [3-7]. Secondly, as a standardized interface between autonomous AI agents and real-world systems, model context protocol (MCP) [8,9] enables scalable and composable development and integration of different kinds of tools for various specific domains.

Therefore, in this demo, as shown in Fig. 1, we propose, implement, and demonstrate an MCP server-based agentic AI architecture, validated using both a simulator/digital twin and real-world testbeds. The proposed solution enables fully automated, vendor-agnostic, multi-layer lifecycle management of IPoDWDM networks, spanning physical layer setup through IP layer provisioning and performance telemetry monitoring.

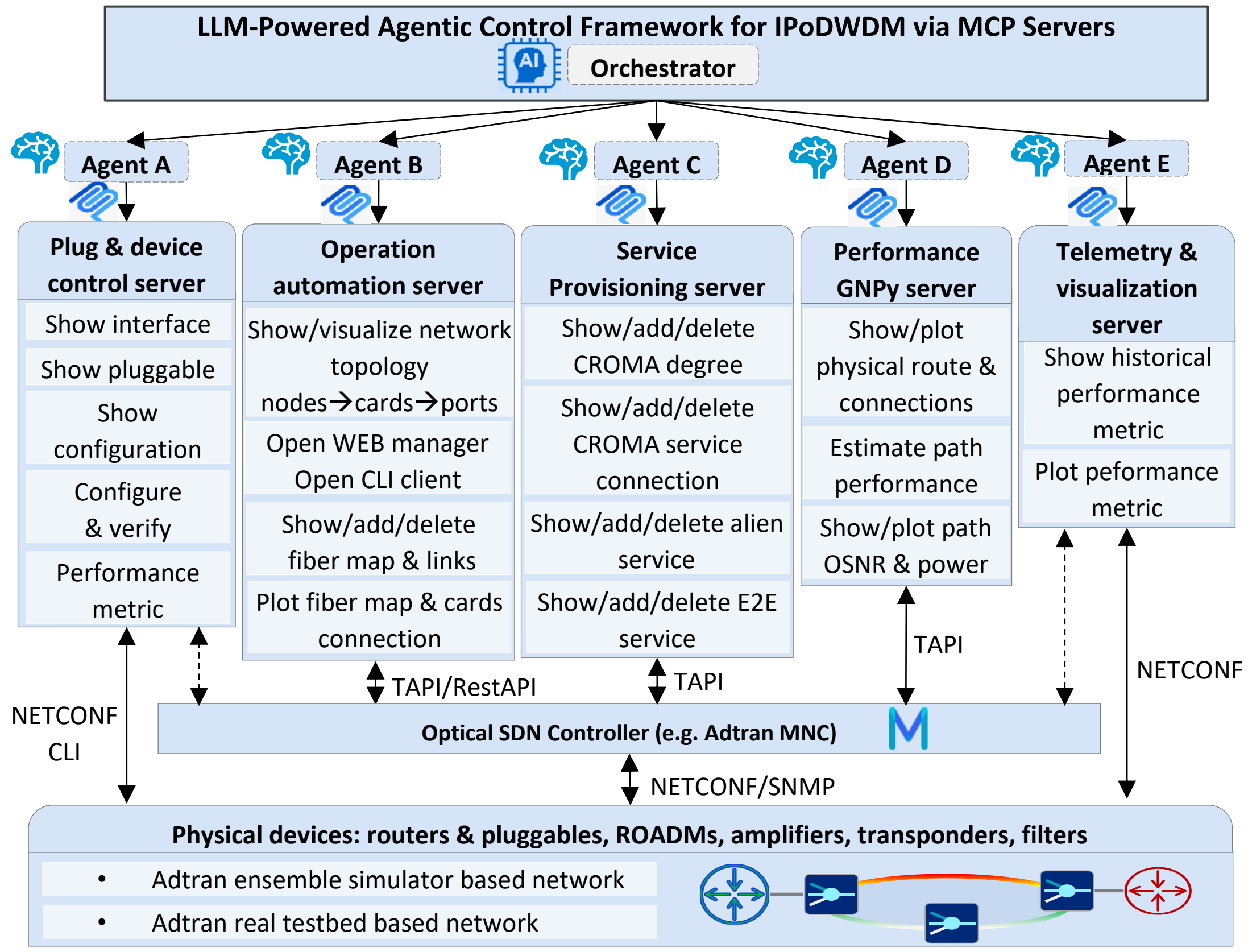

**Fig. 1:** Workflow and architecture of MCP servers based agentic IPoDWDM network

As illustrated in Fig. 1, an LLM (e.g. GPT-5.2) based AI agent coordinates multiple streamable HTTP-based MCP servers, each hosting a collection of MCP tools. These servers interact either directly with network elements, such as routers, or through the Transport API (TAPI) with optical SDN controllers that interface with the underlying network via SNMP or NETCONF. MCP servers are described briefly as follows:

1. ***Plug & device control server***: Check, configure and verify the coherent pluggables of routers or transponders. Performance metrics (e.g. BER and OSNR) are monitored and visualized after service provisioning.
2. ***Operation automation server***: Visualize network topology including detailed nodes, shelves, cards and ports as well as fiber connections. Multiple fiber maps can be created or removed in batch using a single prompt.
3. ***Service provisioning server***: Three MCP tools enable service provisioning via (1) step-by-step ROADM-based configuration by using cloud-based reconfigurable optical management & automation (CROMA), (2) alien wavelength creation at the OLS-router interface, and (3) single-step end-to-end (E2E) provisioning.
4. ***Performance GNPy server***: Estimate and visualize E2E optical performance by translating the discovered network topology and configuration data into a GNPy-compatible JSON model [10].
5. ***Telemetry & Visualization server***: Retrieve and visualize the performance metrics of coherent transceivers from a time series database.

Each MCP server may be controlled by an individual agent, while alternatively a single agent can centrally orchestrate multiple MCP servers.

**Innovation**

As illustrated schematically in Fig. 1, we propose and demonstrate an agentic AI-orchestrated architecture based on multiple MCP servers, enabling full and efficient SDN-based automation from device configuration and logical fiber connectivity to end-to-end service provisioning, performance estimation, and telemetry monitoring. The key novelties are listed but not limited to:

1. Multiple MCP servers covering create, read, update and delete (CRUD) operations cross physical, optical and IP layer to realize the full and deep lifecycle automation.
2. Two user-selectable agentic structures are supported: (1) a single LLM-based agent interfacing with operators and multiple MCP servers; and (2) a hierarchical orchestrator–planner framework coordinating multiple sub-agents (Agent A, B, as shown in Fig. 1). The former offers a unified and responsive interaction model with reduced coordination overhead, while the latter enables the efficient execution of more complex and coordinated tasks.
3. Supporting vendor-agnostic configuration of optical coherent pluggables through abstraction, with automated mapping of Jinja2-based NETCONF templates to device-specific hardware models and OS versions.
4. Rich visualizations integrated with agent and MCP tools: (1) Basic charts (e.g., line or bar) can be generated from numerical data in the response. (2) Advanced interactive visualizations, implemented using a React-based web interface, enhance observability and support efficient port-level operations, including fiber map creation.
5. Stateful interaction is supported by short- and long-term memory for context retention, while tool-calling traces provide full observability of each request, including the agent, server, invoked tools, their arguments, raw outputs, and execution time.

The framework is designed for easy extension: (1) vendors such as Juniper, Cisco, Arista, and Adtran are already supported, and additional vendors can be integrated using NETCONF configurations or CLI commands; (2) new MCP tools, servers, or agents can be added independently alongside existing ones.

**ECOC relevance**

This live demonstration aligns well with the mission and scope of the Demo Zone. Through interactive, natural-language dialogues, attendees can engage with an agent that dynamically invokes multiple MCP servers connected to both digital-twin simulators and a real testbed. Acting as network planners, ECOC participants gain hands-on experience in fully automated workflows, including network setup, fiber connectivity, pluggable configuration, service provisioning, performance prediction, and metrics monitoring, as well as visualizing network topology, physical routes, and card-level connections.

This demo targets or relates to the following ECOC themes:

- AI for optics and optics for AI
- AI/ML for network intelligence, optimization, and predictive control
- Hierarchical packet/optical pluggable control
- Autonomous control, orchestration, and telemetry-based network management

Feedback from ECOC attendees will be used to further refine and validate the proposed approach, supporting its evolution toward practical deployment.

**Demo implementation**

This demo is conducted on-site, primarily showcasing the frontend, specifically the Agent client, to ECOC attendees. The backend components, including the Agent backend, MCP servers, and the Redis database used for storing conversational context, are hosted on remote servers. The MCP servers are remotely connected either directly to devices, such as coherent pluggables in routers, or to the Adtran Mosaic network SDN controller (MNC). The MNC manages either a digital twin–based network or a real physical testbed.

The session begins with a brief introduction to the Agent client GUI, explaining how to input prompts, view the agent tool calling traces and logs, and switch between single-agent and multi-agent modes. Then, all the available MCP servers and tools will be shown to ECOC participants by the prompt “List the tools you see and group them into tables by category”. After that, users can try to show all the existing IPoDWDM networks (simulator based or real testbed) from a given MNC server.

Based on the available tools and exposed network capabilities, users will be guided to perform tasks such as network operations, service provisioning, and performance validation using natural language in an iterative and interactive manner:

- Visualize the network topology, including all elements such as nodes, shelves, cards, and ports. Based on this view, users can create fiber map pairs between any two ports. In addition, internal card-level connections within each node can also be visualized.
- Create or delete multiple fiber maps within a single request.
- Identify available routers and coherent pluggables, and configure then validate selected devices by specifying parameters such as operation mode (or modulation format), frequency, and transmit (Tx) power.
- Provision E2E services between two pluggables using the different approaches described in the “Service Provisioning Server” section of the overview. Users can then verify successful service creation and visualize the associated physical routing elements.
- Predict network performance using the GNPy server. When a real testbed is used, users can also retrieve and visualize actual performance metrics, such as BER and OSNR.

As examples, visualizations of network topology (a), fiber map creations (b), pre-FEC BER (c) and physical routing elements (d) are shown in Fig.2. Additional functionalities and use cases will be demonstrated during the live demo.

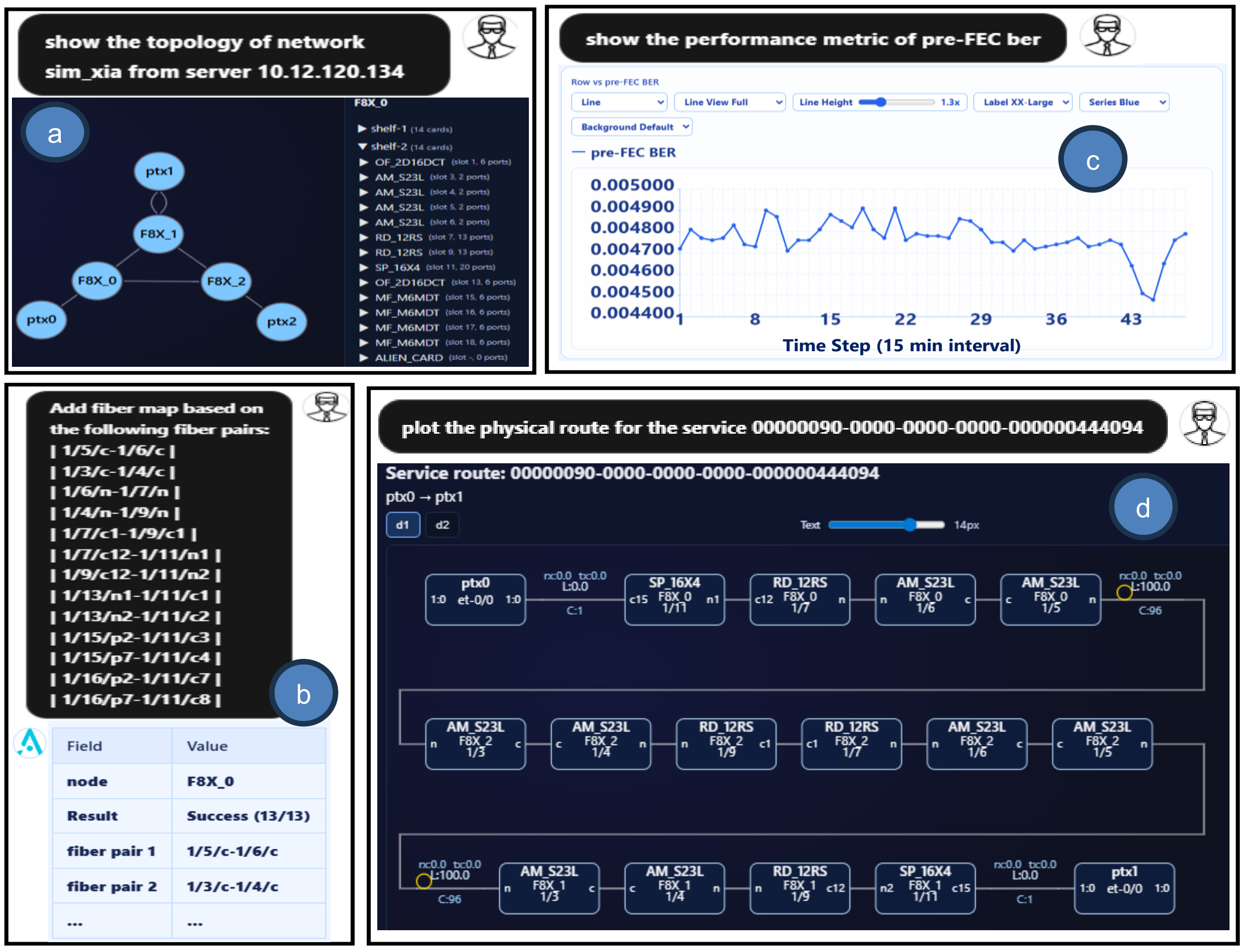

**Fig. 2:** Examples of user requests and responses from the Agent

## Acknowledgment

This work has received funding from the German Federal Ministry of Research, Technology, and Space (BMFTR) project SUSTAINET-Advance, Grant 16KIS2271K, in the framework of the CELTIC-NEXT project id C2024/3-3.